\documentclass[showpacs,preprintnumbers,amsmath,amssymb]{revtex4}

\usepackage{graphicx}
\usepackage{epsfig}
\usepackage{dcolumn}
\usepackage{bm}

\begin{document}

\title {Distribution of the largest fragment in the Lattice Gas Model}

\author{ F.Gulminelli$^{(1)}$ and Ph.Chomaz$^{(2)}$ }

\affiliation{
(1) LPC (IN2P3-CNRS/Ensicaen et Universit\'{e}), F-14050 Caen c\'{e}dex, 
     France \\
(2) GANIL (DSM-CEA/IN2P3-CNRS), B.P.5027, F-14021 Caen c\'{e}dex, France 
 }

\begin{abstract}
The distribution of the largest fragment is studied in different regions
of the Lattice Gas model phase diagram. We show that first and second order 
transitions can be clearly distinguished in the grancanonical ensemble,
while signals typical of a continuous transition are seen inside 
the coexistence region if a mass conservation constraint is applied. 
Some possible implications of these findings for heavy ion multifragmentation
experiments are discussed.
\end{abstract}


\maketitle

\section{Introduction}

Since the first heavy ion experiments the size of the largest cluster $A_{M}$
detected in multifragmentation events has been tentatively
associated to an order parameter for the fragmentation phase
transition\cite{campi}; if this is true, we should expect for this observable
a double humped distribution if the transition is first order\cite{binder},
while its fluctuations should obey the first scaling law if the transition is
continuous\cite{botet}. Experimental multifragmentation data show in this
respect somewhat contradictory evidences. An analysis of 
80 A.MeV Au+Au peripheral collisions from the Indra-Aladin
collaboration\cite{pichon} reports a bimodal distribution of a variable
closely correlated to $A_{M}$.
On the other side the functional relationship between the two first moments
of $A_{M}$ in central Xe+Cu collisions\cite{frankland} 
shows a change of slope which has been
interpreted as a a transition from the $\Delta=1/2$ to the $\Delta=1$ 
scaling law as expected for a generic continuous transition\cite{botet}.
From the theoretical point of view, 
it is well known\cite{dasgupta,bugaev,commande,fisher,raduta} 
that in finite systems 
many different pseudo-critical behaviors 
can be observed inside the coexistence region of a
first order phase transition. In particular concerning the order parameter
fluctuations, simulations have been performed in the framework of the Ising 
Model with Fixed Magnetization (IMFM) in ref.\cite{carmona}. In this study 
the distribution of $A_{M}$ was shown to approximately 
obey the first scaling law 
even at subcritical densities, i.e. in thermodynamic conditions where no 
continuous transition takes place. Since the scaling is violated for very
large lattices, the observed behavior was interpreted in this paper as a finite
size effect that prevents to recognize the order of a transition in a small
system. An important difference subsists though between the theoretical
study of ref.\cite{carmona} and the experimental analysis ref.\cite{frankland}:
in the first paper the average $A_{M}$ size is varied by increasing the 
total lattice size, meaning that the existence of a scaling law is tested 
in well defined thermodynamic conditions (a single point in the ($\rho$,$T$)
state variables space). In the experimental case it is not possible 
to freely  vary the source size, 
therefore different regions of $\langle A_{M}\rangle$
are explored by varying the total energy deposited in the fragmenting system.
It is not a priori clear how these two very different procedures might be related and whether they could be equivalent.  
  
In this paper we analyze the distribution of $A_{M}$ within the Lattice
Gas Model\cite{leeyang}. This model is the simplest representation of the 
liquid-gas phase transition; once augmented with the cluster definition 
through the Coniglio-Klein algorithm\cite{coniglio}, 
it can also be related to a bond and site percolation problem, 
making this model a paradigm of the fragmentation phase transition. 
This model is isomorphous to the Ising spin model 
and its thermodynamic properties are very precisely
known: the Lattice Gas phase diagram contains both first and second order phase
transitions and basic effects, like conservation laws, 
which are very relevant to the experimental
situation, can be easily implemented.

In the analysis of the $A_{M}$ distributions 
we will show that the most important finite size effect is the inequivalence 
between statistical ensembles\cite{inequivalence}: the
observed ambiguities can be coherently interpreted as an
effect of conservation laws, the distribution of an order parameter being 
drastically deformed if a constraint is applied on an observable which is
closely correlated to the order parameter under study. 

Specifically we will show that:
\begin{itemize}
\item in small canonical systems, a first scaling law as a function of the system size can be  observed at the critical
point but also for subcritical densities inside the coexistence region. This is in agreement
with the findings reported in ref.\cite{carmona}. 
The difficulty in recognizing the order of the
transition is not only due to the finite size effects 
but more important, the order parameter 
distribution and its scaling properties are deformed
by the conservation law that in the canonical ensemble acts on the total number of particles $A_{t}$ , 
which strongly constraints the order parameter 
$A_{M}$;
\item if the $A_{M}$ size is varied by changing the system temperature at 
a fixed lattice size, no scaling of the largest fragment distribution 
is observed even if we choose a 
transformation which passes across the thermodynamic critical point;
\item in this case, 
the correlation between the average $< A_M>$ and the variance $ \sigma^2 $ of the largest fragment distribution exhibits a  rise and fall which is imposed by the conservation law constraint; 
the double logarithmic derivative $ \Delta' = d \log \sigma /d < A_M> $ 
appears to be a smooth decreasing function of $< A_M>$; even if 
 $ \Delta'$ is passing through $ \Delta'=1$ and $ \Delta'=1/2$ before 
 becoming negative no simple scalings  
  $ \sigma \propto < A_M> ^\Delta$ can actually be unambiguously
   isolated. 
\item however, we show that 
 both the existence of a transition and a conclusion about its
order can be infered from the quantitative study of the 
$A_{M}$ fluctuation.  
\end{itemize}
 
\section{Phase transition in the Lattice Gas model}

 \begin{figure}[htbp]
\includegraphics*[height=0.4\linewidth]{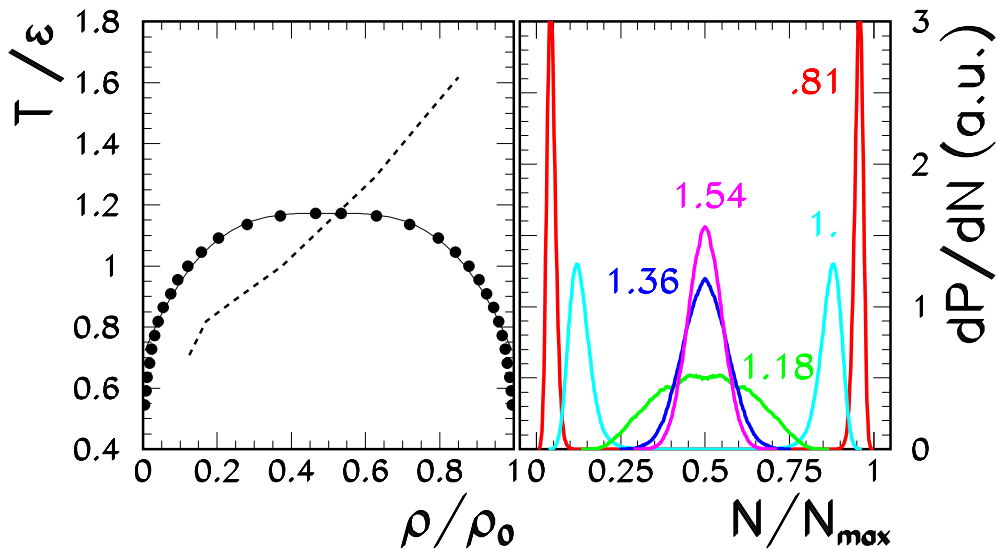}
\caption{Right part: distributions of the size of the largest cluster in the
grancanonical lattice gas model at different temperatures $T/\epsilon$, 
with a 8x8x8 lattice 
and at $\mu=3\epsilon$.
Left side: lattice gas phas diagram from the distributions on the right side.
Dashed line: locus of the maximal $A_{M}$ fluctuation in the canonical 
ensemble.} \label{fig1}
\end{figure}

In our implementation of 
the lattice gas model \cite{leeyang} the $N$ sites  of a cubic
lattice are characterized by an occupation number $n_i$ which is
defined as $n_i =0 (1) $ for a vacancy (particle).
Particles occupying nearest neighboring sites interact with a 
constant coupling $\epsilon$. 
This model can be transformed into an Ising spin problem with a magnetic field
through the mapping $s_i=n_i-1/2$. 
The relative particle density $\rho/\rho_0$ 
is defined as the number of occupied sites
divided by the total number of sites and is linked to the  magnetization of the 
Ising model by $\rho/\rho_0=m+1/2$. 
In addition to this interaction part a kinetic energy is introduced. 
Occupied sites are characterized by a momentum vector.
Observables expectation values are evaluated in the different ensembles
(grancanonical, canonical and microcanonical) sampled
through standard Metropolis algorithms \cite{commande}. The chemical 
potential in the grancanonical implementation plays the role of the magnetic 
field $h=\mu - 3 \epsilon$ in Ising, while the canonical Lattice 
Gas corresponds to the constant magnetization Ising IMFM case with 
$m=\rho/\rho_0-1/2$.

The phase diagram of the model can be easily evaluated looking at the
distribution of the total number of particles $A_{t}=\sum_{i=1}^N n_i$
in the grancanonical ensemble with a chemical potential $\mu=\mu_c=3\epsilon$
which corresponds to the Ising critical field $h=0$.
The $A_{t}$ distributions, $P_{\beta \mu}(A_t)$, are displayed at different 
temperatures in the right 
part of figure \ref{fig1}. The presence of two different ensembles of states (bimodality) is clearly
seen for all temperatures $T<T_c \approx 1.22\epsilon$. 
At the critical chemical potential $\mu_c$
presented in the figure,
the probabilities of occurrence of the two solutions are exactly identical; 
if $\mu<\mu_c$ ($\mu>\mu_c$)
the high (low) density peak dominates.  
For a fixed temperature $\beta^{-1}$, the most probable $A_t$ 
as a function of $\mu$ is discontinuous at the transition point $\mu_c$. 

At the thermodynamic limit, the discontinuity in the most probable $A_t$ as a function of $\mu$ give rise
to a discontinuity in the associated $\langle A_{t} \rangle(\mu)$ equation 
of state;
this implies that the two peaks represent two coexisting phases\cite{zeroes,topology}
and that the number of particles ( or equivalently the density)
is the order parameter of a phase transition which 
is first order up to the critical point $T=T_c$. 
The phase diagram can be constructed by reporting the forbidden region for the most probable density i.e. the locus of the discontinuity in the most probable. This corresponds to the two peaks in the bimodal particle number distribution  observed at $\mu_c$.
The phase diagram is displayed in the left part of figure \ref{fig1}.
These findings obtained in a 8x8x8 lattice 
correspond to the phenomenology of the liquid gas phase
transition that the model is known to display at the thermodynamic limit.
If we increase the lattice size the location of the coexistence border 
will be modified, even if finite size corrections are especially small in this
model\cite{commande}. However it is clear from figure \ref{fig1}
that (except at the critical point 
which is a second order point, where the two peaks
merge to form a single distribution) the first order 
character of the transition is indisputable even for a linear dimension 
as small as $L=8$.

\begin{figure}[htbp]
\includegraphics*[height=0.4\linewidth]{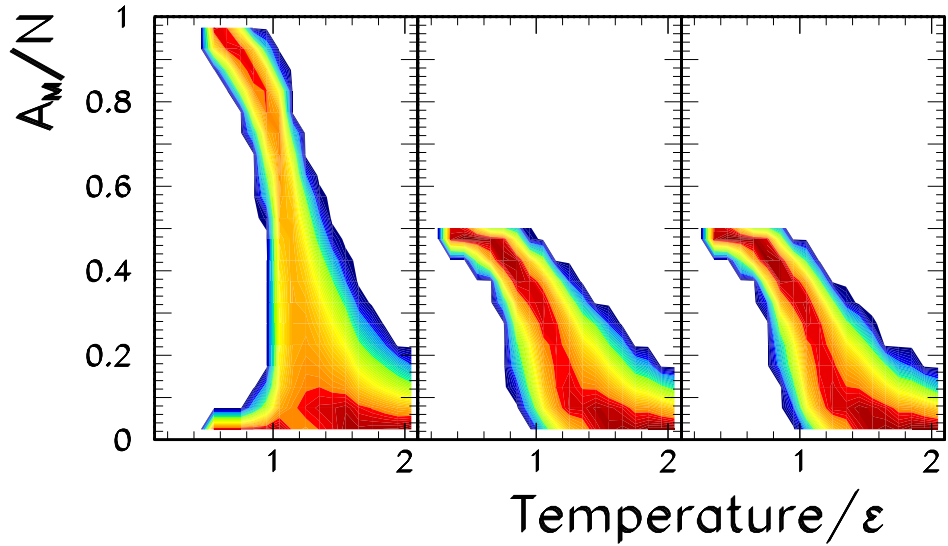}
\caption{ $A_{M}$ distributions as a function of temperature for a 8x8x8
lattice in the
grancanonical (left), canonical (middle), and microcanonical (right) ensemble.
In all cases the density is $\rho/\rho_0=1/2$.} \label{fig2}
\end{figure}

Figure \ref{fig2} shows the size of the largest cluster 
$A_{M}$ as a function of the temperature for the grancanonical, canonical and microcanonical ensembles. The obvious 
correlation between $A_{M}$ and $A_{t}$ 
implies that for  $T<T_c$ the $A_{M}$ distribution 
is also double humped in the grancanonical ensemble 
as explicitly shown in ref.\cite{imfm}. 
This means that $A_{M}$ can also be taken 
as an order parameter of the liquid gas phase transition, and 
looking at its distribution this transition
can be recognized as first order even for a system constituted of 
$\langle A_{t}\rangle = 256$ particles.

\section{Conservation laws and thermodynamics}

If the constraint of mass conservation is implemented (canonical lattice gas,
or equivalently Ising model with fixed magnetization) the distributions  of $A_M$
drastically change\cite{imfm}. In the grancanonical ensemble at 
$\langle \rho/\rho_0 \rangle = 1/2$, the explored
microstates essentially populate the coexistence border, while the coexistence
region is accessed with a negligible probability (see figure \ref{fig1}); 
these 
highly improbable 
of the 
grancanonical distributions are conversely 
the only microstates which are allowed by the
canonical constraint at the value $\rho/\rho_0 = 1/2$
below the transition temperature the grand canonical and canonical 
partitions differ drastically.
Because of the mass conservation constraint, the bimodality of the $A_{t}$
distribution is obviously lost in the canonical ensemble; as a consequence
of the correlation between $A_{t}$ and $A_{M}$,
the $A_{M}$ distribution also shows a unique peak 
(figure \ref{fig2}). 
If we additionally implement a total energy conservation constraint
(microcanonical ensemble, right part of figure \ref{fig2}) 
the distributions get
still narrower, but the qualitative behavior is the same than in the canonical
ensemble.
The normal behavior of the $A_{M}$ distribution at subcritical
temperatures may intuitively suggest a pure phase or a continuous transition for the canonical model.
This intuition is however false; the characteristics
and order of the transition do not depend on the statistical ensemble,
and the phase diagram of figure \ref{fig1} is still pertinent to the canonical
ensemble\cite{imfm}. Indeed the relation between the two ensembles can be 
written as
\begin{equation}
\log P_{\beta \mu} (A_{t}) = \log Z_{\beta} (A_{t}) + \beta \mu A_{t} -
\log Z_{\beta \mu} \label{equiv}
\end{equation}
where $Z_{\beta \mu}$, $Z_{\beta} (A_{t})$ are the partition sums in the two
ensembles.
Eq.(\ref{equiv}) shows that in the whole region where the grancanonical
distribution $P_{\beta \mu} (A_{t})$ is convex, the canonical equation
of state
\begin{equation}
\mu_{can} = - \frac{1}{\beta} \frac{\partial \log Z_{\beta}}{\partial A_{t}}
\label{eos}
\end{equation}
presents a back bending, which is an unambiguous
signal of a first order phase transition\cite{gross}. 
At each temperature 
the maxima of $P_{\beta \mu}$ correspond to the two ending points of the 
tangent construction for eq.(\ref{eos}), i.e. to the borders of the 
coexistence region in the canonical ensemble.

The qualitative behavior of $A_{M}(T)$ in the canonical ensemble does not
change with the density of the system. In particular, 
the $A_{M}$ fluctuation passes systematically through a maximum.
The locus of these maxima is displayed on the phase diagram in 
figure \ref{fig1}. We can see
that the maximum fluctuation approximately corresponds to 
the transition temperature only at the critical point. At subcritical densities 
this maximum lies inside the coexistence region of the first order phase
transition.
The results of figure \ref{fig2} show that the double hump
criterium for a first order phase transition does not hold if a constraint
is put on a variable closely correlated to the order parameter under study.

\section{Conservation laws and Delta scaling}

We can ask the question whether a detailed study of the scaling properties
of the $A_{M}$ distribution may give extra information on the transition
and discriminate first and second order. Following the arguments of
ref.\cite{botet} we consider the distribution
\begin{equation}
\Phi(z) = \Phi\left ( \frac{A_{M}-A_{M}^*}
{\langle A_{M} \rangle^\Delta} \right )
= \langle A_{M} \rangle^\Delta
P(A_{M}) \label{delta_scaling}
\end{equation} 
where $A_{M}^*$ is the most probable value of $A_{M}$ and 
$0<\Delta\leq 1$ is a real number. At a continuous phase transition point
the distribution of the order parameter is expected to fulfil the first
scaling law, i.e. the distribution $\Phi$ should be scale invariant
with $\Delta=1$. The scale invariance of $\Phi$ for a given value of $\Delta$
is generically refered to as $\Delta$ scaling, and the transition
observed experimentally\cite{frankland} from a $\Delta\approx 1/2$ to a 
$\Delta\approx 1$ scaling by varying the centrality of the collision 
and therefore the energy deposited in the system, 
has been taken as a signal of a continuous phase transition.

A practical difficulty in testing $\Delta$ scaling is that for a given
distribution the value of $\Delta$ that corresponds to scale invariance, 
if any,
cannot be known a-priori. This difficulty can be circumvented
using the fact that the scaling (\ref{delta_scaling}) imposed
$<A_M>^\Delta \propto \sigma$.
Then it is immediate to verify that
eq.(\ref{delta_scaling}) can be equivalently written as
the ensemble of the two conditions
\begin{eqnarray}
\Psi\left ( \frac{A_{M}-\widetilde{A}_M}
{ \sigma_{A_{M}} } \right ) &=& \sigma_{A_{M}}  
P(A_{M}) \label{pms} \\
\sigma^2_{A_{M}}&=& K \langle A_{M} \rangle^{2\Delta} \label{cappas}
\end{eqnarray}
where $\Psi$ is a scale invariant distribution and $K$ is a constant.
Since in presence of a scaling, 
the difference between the most probable $A_M^*$ and the average $<A_M>$
scales like $\sigma$, $\widetilde{A}_M$ can either be one or the other.   
In the later case the occurence of  $\Delta$ scaling  study coresponds
 to the invariance of the centered and reduced distribution. If this distribution 
 does not show scale invariance, we can exclude the existence of any 
 $\Delta$ scaling law. 
If the function $\Psi$ is scale invariant, this corresponds to a $\Delta$ 
scaling if and only if the log-log correlation between the average and the
variance is linear; in this case the slope of the correlation gives the value
of $\Delta$. The practical advantage of testing eqs.(\ref{pms},\ref{cappas})
instead of eq.(\ref{delta_scaling}) is that we can check scale invariance
without any a-priori knowledge of $\Delta$.

\begin{figure}[hbtp]
\includegraphics*[height=0.6\linewidth]{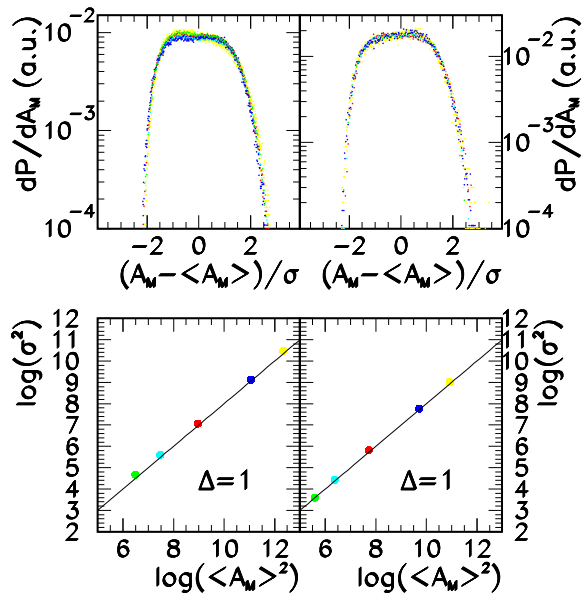}
\caption{Delta scaling analysis for the canonical lattice gas model at the
critical point $\rho=\rho_0/2, T=T_c$ (left side)
and inside coexistence $\rho=\rho_0/4$ at the point of maximal $A_M$
fluctuation (right side). 
Upper part: centered and reduced $A_{M}$ distributions.
Lower part: correlation between the first two moments and linear interpolation
according to eq.(\protect\ref{cappas}).  The linear size of the lattice 
is varied as $L=5,6,8,12,16$. }
\label{fig3}
\end{figure}

The standard way of testing scale invariance is to consider a specific point
of the phase diagram  and 
consider the centered and reduced $A_{M}$ distributions obtained by 
varying the size of the lattice 
and, as a consequence, the total number of particles. 
For the canonical case
at the thermodynamical critical point, 
this analysis is shown on the left side of figure \ref{fig3}. 
Both eq.(\ref{pms}) and eq.(\ref{cappas}) are well
verified, in agreement with the expectation of a first scaling law 
at a continuous transition point\cite{botet}. A comparable quality scaling
is however observed also at subcritical densities
at the temperature corresponding to the maximum $A_{M}$
fluctuations (right side of figure \ref{fig3}). This finding is in agreement with ref.\cite{carmona}. Together with the analysis of the phase diagram
this means that such a scaling also approximately applies
in the coexistence region of a first order phase transition, if the order
parameter is not free to fluctuate but is constrained by a conservation law.

\section{Delta scaling as a function of the system excitation}

\begin{figure}[htbp]
\includegraphics*[height=0.5\linewidth]{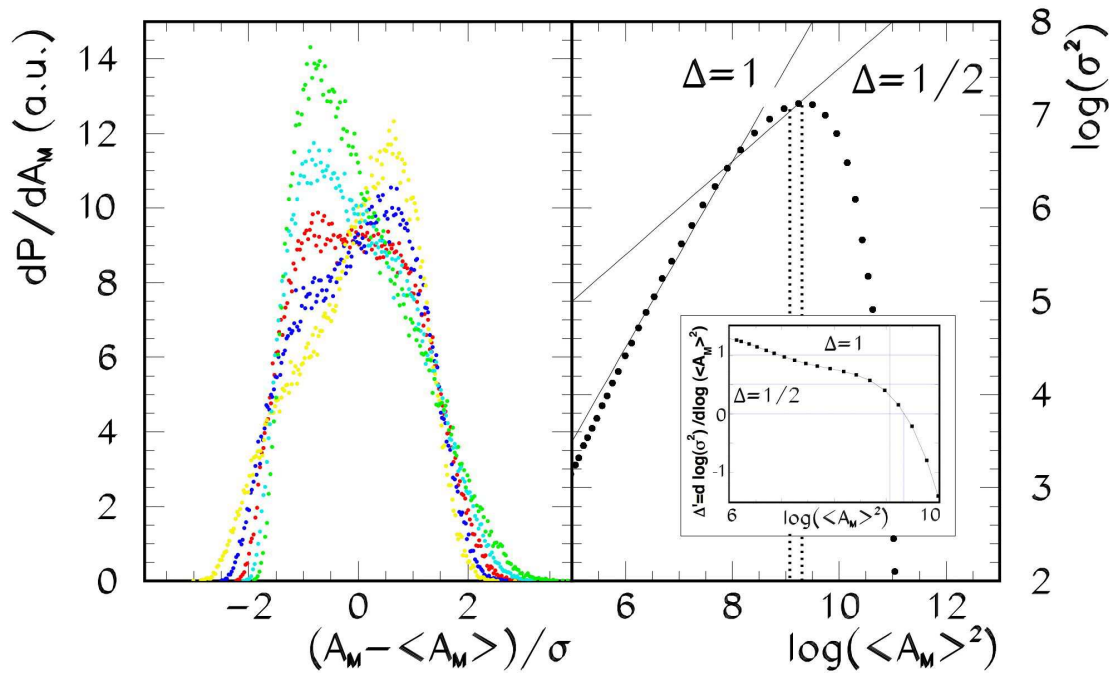}
\caption{Delta scaling analysis for the canonical lattice gas model at 
constant density $\rho=\rho_0/2$ varying the system temperature. 
Left part: centered and reduced $A_{M}$ distributions for temperatures
varying from $T=1.1\epsilon$ to $T=1.25\epsilon$.
Right part: correlation between the first two moments and linear interpolations
according to eq.(\protect\ref{cappas}).  Temperatures range 
from $T=.3\epsilon$ to $T=2.3\epsilon$. The vertical lines indicate the temperature of maximum $A_M$ fluctuations and the critical temperature. The double logarithmic derivative $\Delta'$ is shown in the inserted figure.} \label{fig4}
\end{figure}

In the experimental application to nuclear multifragmentation\cite{frankland}
the system size cannot be varied as freely as in the lattice gas, since the
maximum size for a nuclear system is of the order of 400 particles.
To explore different values of $\langle A_{M} \rangle$, the same system has
been studied at different bombarding energies\cite{frankland} and/or
different impact parameters\cite{frankland_new}. In a similar way, we have kept the total number of particles constant and we have varied the temperature.
To fix the ideas, we have chosen the simplest thermodynamical path 
from coexistence to the fluid phase passing through the critical point, 
$\rho(T)=cte=\rho_0/2$.  
The resulting $\Psi$ functions  are displayed in 
figure \ref{fig4}. No scaling  is observed:
 the function $\Psi$  continuously 
evolves from a  distribution 
with a tail extending towards the low mass side compared to the 
average 
at low temperature while the opposite is true at high temperature.


If we look at the behavior
of the variance as a function of the first moment,
the log-log correlation is nowhere linear showing that 
the large fragment fluctuation does not evolve like a power of the 
average fragment size.
The bell shaped behavior of this curve is due to the mass
conservation constraint, that forces the fluctuation to vanish both 
at low and at high $\langle A_{M} \rangle$ values. 
The observed maximum is in fact the maximum fluctuation point 
shown in figure 1 which, at the critical density, 
occurs close to the critical point and, 
for sub-critical densities, is located inside the coexistence 
region.
   
To qualitatively compare with experimental $\Delta$-scaling analysis, 
we have to remember that 
the studied experimental distributions only cover 
the multifragmentation regime
and do not explore the decreasing part of the 
$\sigma_{A_{M}}(\langle A_{M} \rangle)$ correlation which 
would correspond in the nuclear case to
evaporation from a compound. 
Focusing now on the fragmentation region, 
we show in figure
\ref{fig4} the best power law interpolations of the 
average and variance correlation to be compared 
with the published experimental analysis 
presenting a $\Delta=1$ to a $\Delta=1/2$ regime. 
In this interpretation, the crossing point between 
the two power-law fits  is interpreted as a "transition" 
point. By construction it appears to be at a higher temperature 
then the maximum fluctuation which at this critical density
 comes out to be close to the critical temperature.

To better study the possible occurrence of a power law scaling of the 
large fragment fluctuation we can study the double logarithmic derivative
\begin{eqnarray}
\Delta' &=& \frac{d \sigma_{A_{M}}}{d \langle A_{M} \rangle}  \label{deltaprime}
\end{eqnarray}
In presence of a $\Delta$-scaling this quantity should be constant. Figure  \ref{fig4} 
shows that $\Delta' $ is a smoothly decreasing function passing through the values $1$ and $1/2$ before going through $0$ at the maximum fluctuation point and then becoming negative as a consequence of the mass conservation law. No plateaus of  
$\Delta' $ are observed confirming the absence of scaling.  

%

This violation of scaling occurs in spite of the fact that
a continuous phase transition point (the thermodynamic critical point) is
explored in the simulations. At this point the distributions indeed follow
the first scaling law (left part of figure \ref{fig3}) but this information is
lost if the different distributions are generated by varying the temperature. 
This is not only true for the transition
point, but also for the supercritical regime. Indeed this regime has been shown
to exhibit the second scaling law $\Delta=1/2$ in the Potts model\cite{botet} 
(or something close to it $\Delta\approx 0.6$ for the IMFM\cite{carmona}), while
in the representation of figure \ref{fig4} scaling can everywhere be excluded.

\begin{figure}[htbp]
\includegraphics*[height=0.5\linewidth]{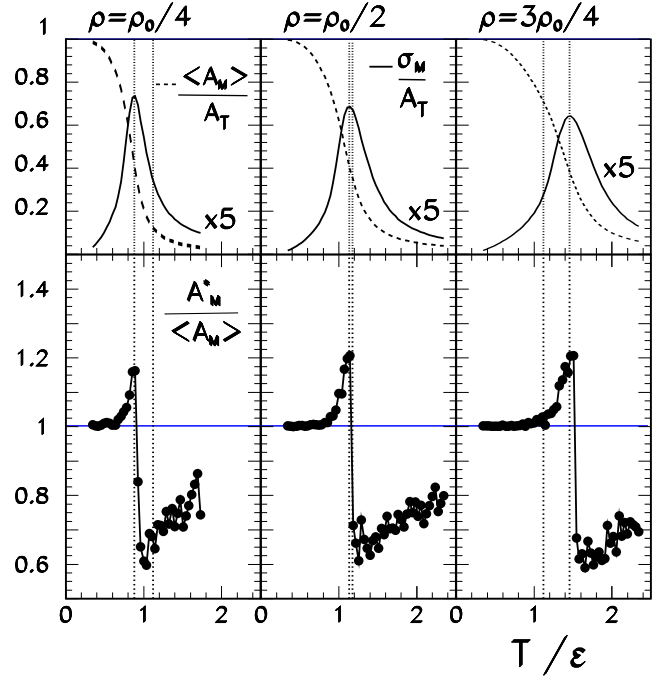}
\caption{ First moments of the $A_M/A_T$ distribution 
as a function of temperature for a 8x8x8 lattice in the canonical ensemble
at $\rho/\rho_0=1/4$ (left),$\rho/\rho_0=1/2$ (medium), and $\rho/\rho_0=3/4$
(right).
Upper part: mean value (full line) and variance (dashed line). 
Lower part: most probable value of the $A_{M}$ distribution
normalized to the mean.  
The vertical lines indicate the temperature of maximal $A_M$ fluctuations 
and the transition temperature for each density.
 }\label{fig5}
\end{figure}

The conclusion is that scale invariance can only be tested 
by varying the total
system size. 
However other information on the
phase transition can be accessed through the study of the
$A_{M}$ distribution with a fixed total number of particles, 
as we now show.
 
\section{Signals of phase transition and of its order}

The first two moments of the distribution in the
canonical ensemble and the corresponding 
most probable value $A_{M}^*$ are displayed in figure \ref{fig5} 
for three different 
densities.
Let us look at the $\rho<\rho_c$ case first.
If the first and second moment show smooth behaviors dominated by the
conservation law constraint, the transition is still apparent in the
behavior of $A_{M}^*$ which rapidly changes at a temperature 
close to the transition point. This sudden decrease is due to a change of sign in the asymmetry of the distribution.
As such, the qualitative behavior of $A_{M}^*(T)$
is independent of the density. 
A great number of continuous transition signals
has been observed in different mass conserving models 
at densities that do not correspond to a continuous phase
transition\cite{carmona,dasgupta,bugaev,commande,fisher,raduta}.
The same happens for the most probable value of $A_{M}$.
This variable shows for all densities 
a sudden drop at a temperature $T_t(\rho)$ which corresponds 
to the maximum of the $A_{M}$ fluctuations. As we have already stressed,
these temperatures approximately coincide with the transition
temperature only at the critical density (see figure \ref{fig1}). 
The behavior at supercritical
densities reflects a geometric phase transition which has no thermodynamic
counterpart, while if fragmentation takes place at low density the $A_{M}$
drop can be taken as a signal of phase coexistence.   

In order to discriminate between the different density regimes 
and recognize the
order of the phase transition, we have to quantify the $A_{M}$ fluctuation
peak. In the grancanonical ensemble, the $A_{t}$ fluctuation is directly
linked to the susceptibility via
\begin{equation}
\chi = \frac{\partial \langle A_{t}\rangle}{\partial \mu}
=\beta (\sigma_{A_{t}}^\mu)^2
\label{macro}
\end{equation} 
To work out a similar expression for the canonical ensemble, let us assume
that $A_{M}$ and the other fragments are statistically independent, i.e. the
total density of states is factorized
\begin{equation}
 W_{t}(A_M,A_m,E_M,E_m)=W_{M}(A_{M},E_M) \cdot W_{m}(A_m,E_m) 
\label{factor}
\end{equation}  
where we have defined the total number of particles not belonging to the
largest fragment as $A_m=A_t-A_M$, and the corresponding energy 
$E_m=E_t-E_M$.
This hypothesis is reasonably well verified in the Lattice Gas model, since
the correlation coefficient between $A_M$ and $A_m$ in the grancanonical
ensemble comes out to be close to zero except in the very dense regime 
$\rho/\rho_0\approx 1$.
The factorization of the state densities implies a convolution 
of the corresponding canonical partition sums
 
\begin{eqnarray}
Z_{\beta}(A_t)&=& \int dE_t e^{-\beta E_t}
\int_0^{E_t}dE_m\int_0^{A_t}dA_m \cdot 
W_m(E_m,A_m) 
W_M(E_t-E_m,A_t-A_m) \nonumber \\
&=& \int_0^{A_{t}} dA_{m} 
Z_\beta^{M}(A_{M})Z_\beta^{m}(A_{t}-A_{M})\label{z_macro}
\end{eqnarray}

where $Z_\beta^{i}, i=m,M$  
describe the contribution of the largest fragment and of all the others
respectively.

The distribution of the largest fragment reads
\begin{equation}
P_{\beta A_{t}} (A_{M}) = Z_\beta^{-1}(A_{t})
Z^M_{\beta}(A_{M})Z^m_{\beta}(A_{t}-A_{M}) 
\label{prob}
\end{equation}

A Gaussian approximation of this distribution leads to\cite{npa}
\begin{equation}
\beta \sigma^2_{A_{M}} = 
\left ( \frac{1}{\chi_m(A_{t}-A_{M}^*)}
- \frac{1}{\chi_M( A_{M}^*)}\right )^{-1}
\label{chineg1}
\end{equation}
where $\sigma^2_{A_{M}}$ is the fluctuation of the $A_M$ distribution
and  the partial susceptibilities  are defined 
as $\chi^{-1}_{i}=\frac{\partial \mu_{i}}{\partial A_{i}}
(A^*_{i})$. 

The above derivation is valid for a system whose state density 
depends on the two extensive variables, number of particles $A$ and energy $E$.
In the case of the fragmentation transition a third extensive variable, the 
volume $V$, has also to be considered. We show in the appendix that in this more general case eq.
(\ref{chineg1}) can still be derived, but a dilute limit 
$V_t=V_m+V_M\approx V_m$ has to be considered.

According to the general definition of phase transitions in finite
systems\cite{gross,houches}, the generalized susceptibility associated to 
an order parameter is negative in a first order phase transition in the
statistical ensemble where the order parameter is subject to a conservation law.
We therefore expect a negative $\chi_M$ at subcritical densities.
Imposing $\chi_M<0$ in eq.(\ref{chineg1}) leads to
\begin{equation}
\sigma^2_{A_{M}} >\beta^{-1}\chi_m(A_{t}- A_{M}^*)
\end{equation} 
Comparing to eq.(\ref{macro}) this finally gives
\begin{equation}
\sigma^2_{A_{M}}=\sigma^2_{A_{m}} >(\sigma^\mu_{A_{m}})^2 
\label{chineg2}
\end{equation}
Equation (\ref{chineg2}) associates the first order phase transition
in the canonical ensemble to "abnormal" $A_{M}$ fluctuations, in the same 
way as abnormal partial energy fluctuations sign a first order phase transition
in the microcanonical ensemble\cite{npa}.
\begin{figure}[htbp]
\includegraphics*[height=0.5\linewidth]{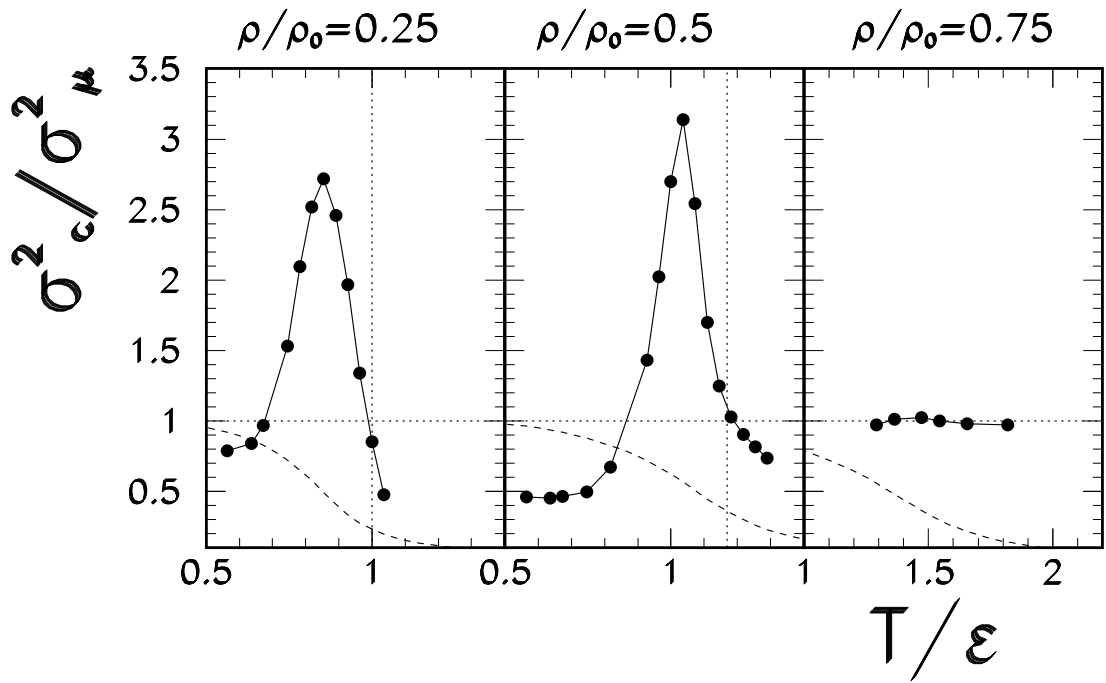}
\caption{Ratio between the grancanonical and canonical 
fluctuation of the number of particles $A_m$ not belonging to the largest
cluster, as a function of the temperature for 
a 8x8x8 lattice  
 at $\rho/\rho_0=1/4$ (left),$\rho/\rho_0=1/2$ (medium), and $\rho/\rho_0=3/4$
(right). Dashed lines: average canonical $A_M$ values normalized to the total 
number of particles.
Vertical lines in the two left panels: limit of 
the region of negative susceptibility from the canonical 
$\mu(A)$ equation of state eq.(\protect\ref{eos})
\protect\cite{commande}. }\label{fig6}
\end{figure}
The canonical and grancanonical fluctuations are compared in figure \ref{fig6}
for three different densities. 
Independent of the system density 
our approximation eq.(\ref{chineg2}) turns out to be incorrect 
at very low temperatures,
when the average size of the largest cluster (dashed lines) exceeds
about 80\% of the available mass. In this case the hypothesis of 
statistical independence between $A_m$ and $A_M$ cannot be justified
and the canonical mass conservation costraint trivially reduces 
the canonical fluctuation. However as soon as the average $A_M$ 
value drops, we can see that the region of negative susceptibility
can be well reconstructed through eq.(\ref{chineg2}),
and in particular its border 
(vertical lines) is very precisely determined by the 
equality condition between the two fluctuations. 
At supercritical densities the dilute gas approximation we have employed
breaks down independent of the temperature 
and the susceptibility cannot quantitatively be estimated
from the fluctuation signal, however in this regime the relative fluctuation
observable does not present any peak while
only inside the spinodal region of the first order phase transition
the canonical fluctuation exceeds the grancanonical one. 
It is clear that this observable allows a unambiguous discrimination 
between the supercritical regime and phase coexistence.
 
\section{Conclusions}

To conclude, in this paper we have discussed the role of the largest fragment in
the framework of the lattice gas model. We have shown that this variable can be
taken as an order parameter of the fragmentation phase transition if this latter
belongs to the liquid gas universality class. It has been already observed
\cite{carmona,pliemling} that the phase transition can be tracked from the
sudden drop of $A_{M}$ close to the transition temperature. This drop is well
fitted by a power law with a $\beta$ exponent close to the expected value for
the liquid gas universality class\cite{pliemling} but finite size effects blur
the behavior considerably for system sizes comparable to accessible nuclear
sizes. 
However, when no constraints are affecting the fluctuations of the 
order parameter such as in the grand canonical ensemble,
 we have shown that the transition is very well defined
if instead of the average we look at the most probable value of $A_{M}$.
Indeed crossing a first order phase transition point this variable is
discontinuous independent of the system size. 
The important result is that if we look at this variable finite
size effects do not constitute a major problem to identify a phase transition
nor to recognize its order. 

On the other hand important ambiguities arise from the non equivalence of
statistical ensembles inside a phase transition. Indeed the distribution of the
order parameter is strongly deformed by the presence of conservation laws in the
system under study. If we look at $A_{M}$ as an order parameter, 
the double hump criterium for a first order phase transition
does not apply any more in the canonical or microcanonical ensemble because of
the strong correlation between the conserved total number of particles and the
order parameter. The mass conservation constraint induces a maximum in the
fluctuation of $A_{M}$ which is not necessarily correlated with the properties
of the phase diagram. We observe maxima both at the critical density close to the 
critical point and at sub-critical densities inside the coexistence zone. 
Moreover,  
the presence of this maximum can simulate a transition
from a $\Delta=1$ to a $\Delta=1/2$ scaling law in a region above the maximum 
fluctuation. 
It is clear that other
observables have to be employed if we want to conclude about the order and
nature of the phase transition. 
One such observable is the numerical value 
of the fluctuation of $A_{M}$, which is by construction identical to the
fluctuation of the number of particles that do not belong to the largest 
cluster $A_m$: if and only if the system crosses the phase coexistence region of
a first order phase transition, this fluctuation overcomes the corresponding
value in the grancanonical ensemble. 

\section{Appendix: derivation of eq.(10)}

The density of states is a function of all the relevant extensive variables
of the system. For the lattice gas model this means $W=W(E,A,V)$.
If the largest fragment $A_{M}$  is statistically independent 
from the other clusters then 
\begin{equation}
 W_{t}(A_M,E_{M},V_M,A_m,E_m,V_m)=W_{M}(A_{M},E_{M},V_M) \cdot 
 W_{m}(A_m,E_m,V_m) 
\end{equation}  
where we have defined the total number of particles not belonging to the
largest fragment as $A_m=A_t-A_M$, and the corresponding energy and volume
$E_m=E_t-E_M$,$V_m=V_t-V_M$.
Let us first consider the case of an external temperature $T=\beta^{-1}$ 
and pressure $p=\beta\lambda$.
Using the standard definition of the canonical isobar partition sum
\[
Z_{\beta\lambda}(A)=\int dE e^{-\beta E}\int dV e^{-pV}W(E,A,V)
\label{zeta_partial}
\]
the total partition sum can be written as
\begin{eqnarray}
Z_{\beta\lambda}(A_t)&=&\int dE_t e^{-\beta E_t}\int dV_t e^{-pV_t}
\int_0^{E_t}dE_m\int_0^{V_t}dV_m\int_0^{A_t}dA_m \cdot \nonumber \\
&&W_m(E_m,A_m,V_m) 
W_M(E_t-E_m,A_t-A_m,V_t-V_m) \nonumber
\end{eqnarray}
or equivalently
\begin{equation}
Z_\beta(A_{t}) = \int_0^{A_{t}} dA_{m} 
Z_\beta^{M}(A_{M})Z_\beta^{m}(A_{t}-A_{M})
\label{z_macro2}
\end{equation} 
where $Z_\beta^{i}, i=m,M$  
describe the contribution of the largest fragment and of all the others
respectively.
In the isochore case $V_m+V_M$=cte, the convolution of the partition
sum is less straightforward because of the presence of the volume
integral
\begin{equation}
Z_\beta(A_{t},V_t) = \int_0^{A_{t}} dA_{m} \int_0^{V_{t}} dV_{m}
Z_\beta^{M}(A_{M},V_M)Z_\beta^{m}(A_{t}-A_{M},V_t-V_M)
\end{equation}
Let us introduce the partial pressures 
$p_{i}=\beta^{-1}\frac{\partial ln Z_\beta^{i}}{\partial V_{i}}
(A^*_{i},V^*_{i})$ and chemical potentials
$\mu_{i}=\beta^{-1}\frac{\partial ln Z_\beta^{i}}{\partial A_{i}}
(A^*_{i},V^*_{i})$ at the most probable volume and mass partition
$A^*_M,V^*_M$. Equilibrium between the two components 
implies $\mu_m=\mu_M$, $p_m=p_M$.  
A saddle point approximation then gives
\begin{eqnarray}
Z_{\beta}^{M} Z_{\beta}^{m} \approx
&exp& \left( -\beta \left[ A^*_M f_M + A^*_m f_m 
\right] \right) \cdot
\nonumber \\
&exp& \left( -\beta \left[ \frac{1}{2}
\left( A_m-A^*_m\right)^2 \left( \chi_M^{-1}+ \chi_m^{-1}\right)
+ \frac{1}{2}
\left( V_m-V^*_m\right) ^2 \left( \kappa_M^{-1}+ \kappa_m^{-1}\right)
\right] \right) \cdot
\nonumber \\
&exp& \left( -\beta \left[
\frac{1}{2}
\left( A_m-A^*_m\right) \left( V_m-V^*_m\right) 
\left( \frac{\partial p_M}{\partial A_M}
+ \frac{\partial p_m}{\partial A_m}
\right) \right] \right) 
\nonumber 
\end{eqnarray}
where $f_i=-T lnZ_{\beta}^{i}(A^*_i)/A^*_i, i=m,M$ are the most 
probable free energies per particle,  
the partial susceptibilities and compressibilities are defined 
as $\chi^{-1}_{i}=\frac{\partial \mu_{i}}{\partial A_{i}}
(A^*_{i},V^*_{i})$, 
$\kappa^{-1}_{i}=\frac{\partial p_{i}}{\partial V_{i}}
(A^*_{i},V^*_{i})$, and the conservation constraints make the linear
terms vanish.
In the dilute limit $V_t=V_m+V_M\approx V_m$ the density variation of the 
"gas" component $m$ is due to its number variation 
$d\rho_m=dA_m/V_t$ and the volume variation can be neglected respect 
to the number variation $V_m-V^*_m \ll A_m-A^*_m$
giving
\begin{equation}
Z_\beta(A_{t},V_t) \approx \int_0^{A_{t}} dA_{m} 
Z_\beta^{M}(A_{M},V^*_m)Z_\beta^{(m)}(A_{t}-A_{M},V_t-V^*_m)
\label{z_cano}
\end{equation} 

Both in the isobar (\ref{z_macro2}) and in the isochore (\ref{z_cano}) 
case, the distribution of the largest fragment reads
\begin{equation}
P_{\beta A_{t}} (A_{M}) = Z_\beta^{-1}
Z^M_{\beta}(A_{M})Z^m_{\beta}(A_{t}-A_{M}) 
\end{equation}
Implementing the saddle point approximation we can identify
\begin{equation}
\beta \sigma^2_{A_{M}} = 
\left ( \frac{1}{\chi_m(A_{t}-A_{M}^*)}
- \frac{1}{\chi_M( A_{M}^*)}\right )^{-1}
\end{equation}
where $\sigma^2_{A_{M}}$ is the fluctuation of the $A_M$ distribution.


\begin{thebibliography}{99}

\bibitem{campi} X. Campi, J.Desbois, E.Lipparini, 
Phys. Lett. 138B (1984) 353.
\bibitem{binder} K. Binder, D. P. Landau, Phys. Rev. B 30 (1984) 1477.
\bibitem{botet} R. Botet, M. Ploszajczak, Phys. Rev. E62 (2000) 1825.
\bibitem{pichon} B. Tamain et al., Nucl.Phys.A, in press;
M.F.Rivet et al., nucl-ex/0412007.  
\bibitem{frankland}R.Botet et al., Phys.Rev.Lett.86(2001)3514.
\bibitem{dasgupta} C. B. Das, S. Das Gupta and A. Majumder, Phys. Rev. C65
(2002) 34608; C. B. Das et al., Phys. Rev. C66 (2002) 044602.
\bibitem{bugaev} K.~A.~Bugaev et al., Phys. Rev. C62 (2000) 044320
                and Phys. Lett. B 498 (2001) 144.
\bibitem{commande} F.Gulminelli, Ph.Chomaz, Phys.Rev.Lett.82 (1999) 1402;
Ph.Chomaz, F.Gulminelli, Int. Journ. Mod. Phys. E8 (1999) 527.
\bibitem{fisher} F. Gulminelli et al., Phys. Rev.C (2002) 51601.
\bibitem{raduta} A. H. Raduta et al., Phys. Rev. C 65, 034606 (2002).
\bibitem{carmona} J.~M.~Carmona, J.~Richert, P.~Wagner, Phys. Lett.
B531 (2002) 71.
\bibitem{leeyang} {C.N.Yang, Phys.Rev.85 (1952)809}
\bibitem{coniglio}  A. Coniglio and W. Klein, J. Phys. A13 (1980) 2775;
X. Campi, H. Krivine and A. Puente, Physica A 262 (1999) 328.
\bibitem{inequivalence} F. Gulminelli and Ph. Chomaz, 
Phys.Rev.E 66 (2002) 046108. 
\bibitem{zeroes} K. C. Lee, Phys. Rev. 53 E (1996) 6558;
Ph. Chomaz, F. Gulminelli, Physica A (2003).
\bibitem{topology} Ph.Chomaz, F.Gulminelli, V.Duflot, Phys.Rev.E64 (2001)
                        046114.
\bibitem{imfm} F. Gulminelli et al.,  Phys. Rev. E 68 (2003) 026120.
\bibitem{gross} D. H. E. Gross,  ''Microcanonical
thermodynamics: phase transitions in finite systems'', Lecture notes in
Physics vol. 66, World Scientific (2001).
\bibitem{npa} P.Chomaz and F.Gulminelli, Nucl.~Phys. A647 (1999) 153.
\bibitem{houches}Ph.Chomaz and F.Gulminelli, in  
'Dynamics and Thermodynamics of systems with long range interactions', 
Lecture Notes in Physics vol.602, Springer (2002).
\bibitem{frankland_new}J.Frankland et al., nucl-ex/0404024
unpublished.
\bibitem{pliemling}  M. Pleimling and A. Hueller, 
J. Stat. Phys. 104 (2001) 971.

\end{thebibliography}
\end{document}